\documentclass[prl,twocolumn,superscriptaddress,amsmath,amssymb]{revtex4}
\usepackage{graphicx}
\usepackage{dcolumn}
\usepackage{bm}
\usepackage{color}

\begin{document}

\title{Non-thermal hot electrons ultrafastly generating
hot optical phonons in graphite}

\author{Y.~Ishida}
\affiliation{ISSP, University of Tokyo,
Kashiwa, Chiba 277-8581, Japan}
\affiliation{RIKEN SPring-8 Center, Sayo, Hyogo 679-5148, Japan}

\author{T.~Togashi}
\affiliation{RIKEN SPring-8 Center, Sayo, Hyogo 679-5148, Japan}

\author{K.~Yamamoto}
\affiliation{RIKEN SPring-8 Center, Sayo, Hyogo 679-5148, Japan}
\affiliation{Graduate School of Engineering, Osaka Prefecture University, 
Sakai, Osaka 599-8531, Japan}

\author{M.~Tanaka}
\author{T.~Taniuchi}
\author{T.~Kiss}
\author{M.~Nakajima}
\author{T.~Suemoto}
\affiliation{ISSP, University of Tokyo,
Kashiwa, Chiba 277-8581, Japan}

\author{S.~Shin}
\affiliation{ISSP, University of Tokyo,
Kashiwa, Chiba 277-8581, Japan}
\affiliation{RIKEN SPring-8 Center, Sayo, Hyogo 679-5148, Japan}
\affiliation{CREST, Japan Science and Technology Agency, 
Tokyo 102-0075, Japan}

\begin{abstract}
Ultrafast dynamics of graphite is investigated by 
time-resolved photoemission spectroscopy. We observe spectral 
features of direct photoexcitations, non-thermal electron distributions, 
and recovery dynamics occurring with two time scales having distinct 
pump-power dependences. Additionally, we find an anomalous increase 
of the spectral intensity around the Fermi level, 
and we attribute this to spectral broadenings due to 
coupled optical phonons in the transient. 
The fingerprints of the coupled optical phonons occur 
from the temporal region where the electronic temperature 
is still not definable. This implies that there is a mechanism 
of ultrafast-and-efficient phonon generations beyond a 
two-temperature model.
\end{abstract}

\maketitle

\textbf{
Investigation of the non-equilibrium dynamics after an impulsive impact 
provides insights into couplings among various excitations. 
A two-temperature model 
(TTM) is often a starting point to understand the coupled dynamics 
of electrons and lattice vibrations: the optical pulse primarily 
raises the electronic temperature $T_{\it el}$ 
while leaving the lattice temperature $T_l$ low; subsequently 
the hot electrons heat up the lattice until $T_{\it el} = T_l$ 
is reached. This temporal hierarchy owes to the assumption that 
the electron-electron scattering rate is much larger than 
the electron-phonon scattering rate. 
We report herein that the TTM scheme is seriously 
invalidated in semimetal graphite. Time-resolved 
photoemission spectroscopy (TrPES) of graphite reveals that 
fingerprints of coupled optical phonons (COPs) occur 
from the initial moments 
where $T_{\it el}$ is still not definable. 
Our study shows that ultrafast-and-efficient phonon generations 
occur beyond the TTM scheme, presumably associated  
to the long duration of the non-thermal electrons in graphite.
}

\begin{figure}
\begin{center}
\includegraphics[width=84mm]{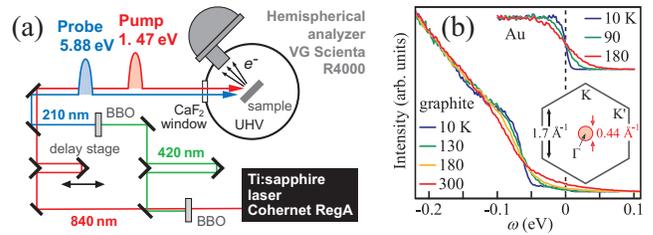}
\end{center}
\caption{\label{fig1}TrPES system. 
(a) A schematic of the TrPES system. 
(b) Spectra of graphite recorded in normal-emission geometry. 
The increase of the intensity at $E_F$ with increasing $T$ 
is due to the increased population of the COPs 
\cite{Liu_2010PRL}.
Inset of (b) shows the 
surface Brillouin zone of graphite, and the area probed 
in the normal-emission geometry 
is indicated by a circle. }
\end{figure}

One of the most interesting observation in the ultrafast dynamics 
of graphite is that $T_{\it el}$ in the sub-picosecond temporal 
region stays somewhat low even though irradiated with an intense 
femtosecond optical pulse \cite{Krampfrath}. This led to the picture 
that the electronic energy is quasi-instantaneously transferred 
to the COPs through strong electron-phonon couplings \cite{Krampfrath} 
based on the TTM scheme \cite{Allen, Brorson, Perfetti_Bi2212}. 
However, electron-phonon coupling constant is reported to be moderately small 
in graphite \cite{Leem_ARPES}, making the mechanism of the 
ultrafast COP generation elusive. Subsequent studies also suggest 
the nearly instantaneous COP generation coupled to the electron dynamics 
\cite{Dawlaty, Ishioka, SDong, Hertel_nanotube, Heinz_Stiff, 
Breusing, Spencer, Kang}, which is considered to affect 
ballistic transports at high fields \cite{Yao, NanoLett10, 
Hwang-Hu-Sarma, Tse-Hwang-Sarma, Calandra-Mauri}. 
Nevertheless, direct observation of the electron distribution 
in the transient is limited \cite{SDong, Breusing, Xu, Moos}, 
and moreover, simultaneous detection of the electron distribution 
and the phonons in the transient has been beyond reach. 
TrPES [Fig.\ \ref{fig1}(a)] 
is one of the most powerful tools to investigate the dynamics of 
the electrons, since it can provide information of the transient 
electron distributions in a wide energy range across the Fermi level ($E_F$) 
\cite{Xu, Moos, Bokor, Perifetti_TaS2, Perfetti_Bi2212, TbTe3}. 
Furthermore, Liu {\it et al.}\ \cite{Liu_2010PRL} recently reported that 
fingerprints of COPs occur in the photoemission spectra of graphite, 
as we shall explain below. Therefore, it became possible to monitor 
simultaneously the electron distribution and the fingerprints of 
COPs during the ultrafast dynamics of graphite by TrPES.

\begin{figure*}
\begin{center}
\includegraphics[width=150mm]{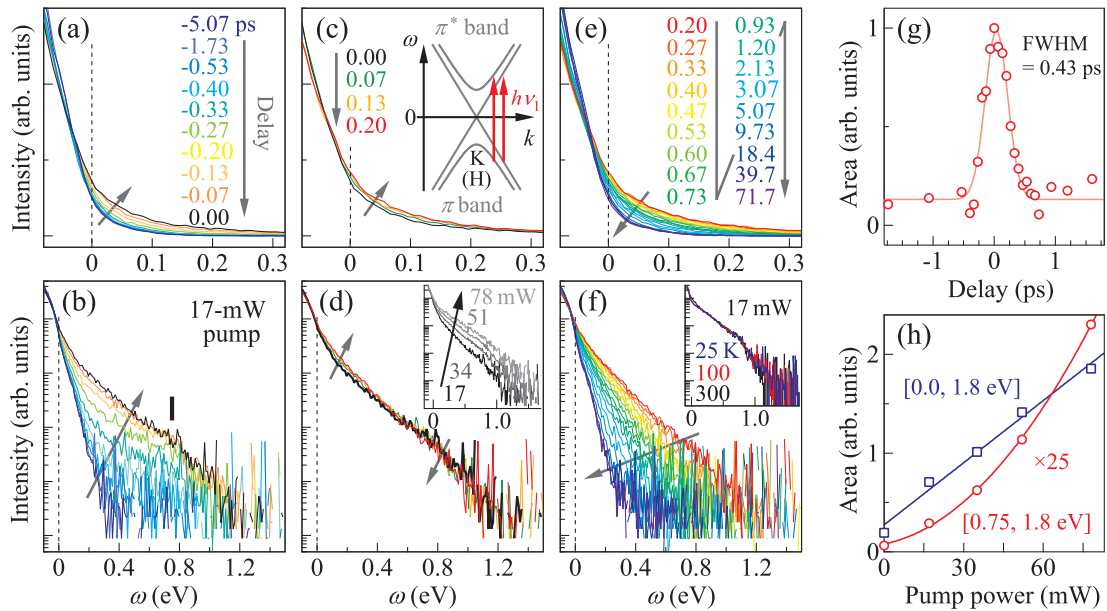}
\end{center}
\caption{\label{fig2}TrPES of graphite
Spectra recorded at 17-mW pump at 
$t\leq 0$ ps (a, b), 0 $\leq t\leq 0.2$ ps (c, d), and 
$t\geq$ 0.2 ps (e, f). 
Here, (b, d, f) are semi-logarithmic plots of (a, b, c), respectively. 
Insets in (d) and (f) show spectra at $t$ = 0 ps 
recorded at various pump powers and temperatures, respectively. 
(g) Spectral weight between 0.8\,-\,1.0 eV 
as a function of $t$ overlaid with a Gaussian. 
(h) Spectral weight under 0.75\,-\,1.8 eV (above the cutoff; circle) 
and under 0.0\,-\,1.8 eV (in the unoccupied side; square) 
at $t$ = 0 ps as a function of 
pump power overlaid with first and second order 
polynomial functions, respectively. 
The schematic in (c) shows direct excitations from the 
$\pi$ bands to the $\pi^*$ bands.}
\end{figure*}

The fingerprints of COPs show up in the spectra recorded 
in a normal-emission geometry \cite{Liu_2010PRL}, that is, when we detect 
the photoelectrons around $\Gamma$ of the surface Brillouin 
zone of graphite [inset in Fig.\ \ref{fig1}(b)]. 
Since there are no bands around $\Gamma$ in the vicinity of $E_F$, 
the signal consists of photoelectrons around $K$ ($K'$) 
indirectly scattered 
into the vicinity of $\Gamma$ mainly by 
phonons. 
As we shall see later, signals of direct photoexcitations 
around the $K$ ($K'$) point indeed occur in the TrPES spectra 
recorded in the normal-emission geometry. 
A gap-like feature of $\sim$70\,meV occurs 
in the spectra near-$E_F$ at low $T$ [Fig.\ \ref{fig1}(b)], 
since phonon absorption process is quenched and phonon emission 
(energy-loss) process dominates. 
With increasing $T$, 
phonon absorption process is increased, 
and the spectral weight tails into higher energies resulting in 
an increase of the spectral 
intensity at $E_F$ [Fig.\ \ref{fig1}(b)]. 
This is similar to anti-Stokes lines in Raman spectra 
gaining stronger intensity at higher temperatures. 
Thus, we utilize the spectral intensity at $E_F$ in TrPES as a measure 
of the number of COPs in the transient. 
The optical phonons monitored herein are assigned to the 
67-meV out-of-plane and the $\sim$150-meV in-plane modes \cite{Liu_2010PRL}.

{\bf Results}.--- 
Figure \ref{fig2}(a-f) show TrPES spectra of graphite, $I(\omega, t)$, 
recorded under a pump power $p$\,$=$\,17 mW 
(a fluence of 
$\sim$14 $\mu$J/cm$^{-2}$) at room temperature. 
A movie file is provided in Supplementary Information. 
Overall, we observe that the electrons are 
pumped from the occupied side to the unoccupied side 
and subsequent recovery dynamics lasting over several tens of 
picoseconds: 
at $t$\,$\le$\,0 ps, 
the spectral intensity $I(\omega, t)$ is increased 
in the unoccupied side; 
at 0\,$\le$\,$t$\,$\le$\,0.20 ps, $I(\omega, t)$ 
at $\omega$\,$\sim$\,0.75 eV is decreased, 
whereas that at $\omega$\,$\sim$\,0 eV is 
increased; 
at $t$\,$\ge$\,0.20 ps, $I(\omega, t)$ in the 
unoccupied side is decreased. Here, 
$t$\,=\,0 ps has been determined utilizing the 
fast response of $\sim$20 fs observed at $\omega >$ 0.8 eV 
\cite{Breusing}, 
and the time resolution 
[full width at half the maximum (FWHM)] 
is estimated to be 
$\varDelta t$\,=\,0.43\,ps, see Fig.\ \ref{fig2}(g).

We observe a plateau feature in the unoccupied side 
during the pump [Fig.\ \ref{fig2}(b) and 
the movie file in Supplementary Information], 
which is a hallmark of a non-thermal electron distribution \cite{Bokor}. 
The plateau turns over into an exponential tailing 
at $t$\,$\gtrsim$\,0.20 ps [Fig.\ \ref{fig2}(f)], 
indicating that the 
electrons are distributed according to the Fermi-Dirac function. 
That is, electronic thermalization occurs at $\tau_e$\,$\sim$\,0.2 ps 
so that $T_{\it el}$ becomes definable thereafter. 
In metallic materials, typical time scale for electronic thermalization 
is considered to be $\tau_e$\,$\sim$\,10\,fs 
or less \cite{Allen}, and if $\varDelta t$\,$\gg$\,$\tau_e$, 
one would not expect to observe a non-thermal distribution of the 
hot electrons. In fact, in the 
TrPES study of a metallic Bi$_2$Sr$_2$CaCu$_2$O$_{8+\delta}$ 
\cite{Perfetti_Bi2212}, 
Perfetti {\it et al}.\ observed with $\varDelta t$\,$\sim$\,90\,fs 
that the spectra mostly obey Fermi-Dirac statistics even at 
$t$\,$\sim$\,0\,ps. 
The smallness of 1/$\tau_e$ in graphite 
can be attributed to  
the semimetallic 
band structure: Since electronic states near $E_F$ 
occur only around $K$($K'$) points of the surface 
Brillouin zone, the available phase space 
for electron-electron scatterings becomes vanishingly small 
in approaching $E_F$ particularly after the initial avalanche 
of the hot electrons towards $E_F$. This can act as a bottleneck for the 
non-thermal electrons to relax into the 
thermal (Fermi-Dirac) distribution.

The plateau observed just after the pump 
extends up to 
a cutoff at $\omega$\,$\sim$\,0.75\,eV, as 
indicated by a bar in Fig.\ \ref{fig2}(b). 
This cutoff can be understood as a fingerprint of 
direct photoexcitations occurring around the $K$($K'$) point: 
since the $\pi$- and $\pi^*$-bands are 
nearly symmetric about $E_F$ 
as shown in the schematic in Fig.\ \ref{fig2}, 
direct excitations are dominated by the transitions 
from $\omega= -$$h\nu_1/2$ 
to +$h\nu_1/2$, 
and therefore, the cutoff energy can be identified to 
$\omega$\,$\sim$\,$h\nu_1$/2\,=\,0.75 eV. 
Note that such a fingerprint of direct excitations would not 
be detected if $\tau_e$\,$\ll$\,$\varDelta t$. Therefore the 
$\sim$0.75-eV cutoff strengthens the conclusion that the 
duration of the non-thermal electron distribution is detected 
in the present study.

With increasing $p$, the cutoff is blurred 
[inset in Fig.\ \ref{fig2}(d)], so that 
the unoccupied side of the spectra becomes 
featureless and the line shape becomes similar to 
the TrPES spectra of graphite reported previously \cite{Moos}. 
The intensity above the cutoff at $t$ = 0 ps grows quadratically with $p$ 
while the intensity in the unoccupied side 
grows linearly with $p$ [Fig.\ \ref{fig2}(h)]. 
The latter indicates that the number of electrons excited by the 
pump is proportional to $p$ at least up to 78 mW, 
whereas the former indicates 
that the excited electrons are further scattered above the cutoff 
by other excitations such as the hot electrons themselves or 
hot phonons, since such scatterings 
occur roughly proportional to the square of the 
population of the excited particles. 
The intensity and the sharpness of the 
cutoff are almost independent of $T$ 
[inset of Fig.\ \ref{fig2}(f)], indicating 
that the blurring of the cutoff is not related to the 
thermally populated excitations in the initial state.

We now turn to the variation of the spectral weight 
around $E_F$, which serves as the fingerprints of COPs in the 
transient, as explained previously. 
One can see that the variation starts from the beginning of the transient 
in accord with the reports of the nearly instantaneous generation of 
the COPs in graphite \cite{Ishioka, Heinz_Stiff, Kang}. 
Our findings therefore show that the ultrafast COP generation takes place 
from the temporal region where the hot electrons are 
not thermalized 
[see, Fig.\ \ref{fig2}(a) and the movie file in Supplementary Information]. 
After $t \sim$ 0.2 ps, the intensity at $E_F$ starts to decrease, 
see Fig.\ \ref{fig2}(e), 
indicating that the COP generation is mosltly accomplished 
within $t \lesssim$ 0.2 ps. The results are in strong contrast 
to the TTM scheme, where COPs are assumed to be cool at the 
beginning and then gradually heated up by the electrons 
that are thermalized.

\begin{figure}
\begin{center}
\includegraphics[width=87mm]{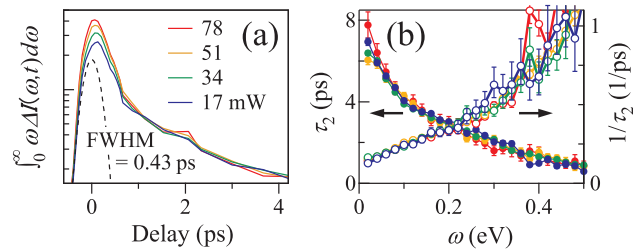}
\end{center}
\caption{\label{fig3}Decay rates. 
(a) Electronic energy (see text) {\it vs} 
$t$ for various pump powers. 
Each curve has an arbitrary offset. 
Dashed line represents the time resolution. 
(b) The spectra of decay time and decay rate 
(left and right axes, respectively) of the slow component. }
\end{figure}

{\bf Analysis}.---
First, we investigate how the total electronic energy dissipates 
with time. As $I(\omega, t)$ reflects the occupied density of states 
(DOS), $\int_{0}^{\infty}\omega\varDelta I(\omega, t)\,d\omega$ 
is a measure of the excess electronic energy, and we plot this as 
a function of $t$ in Fig.\ \ref{fig3}(a). Here, 
$\varDelta I(\omega, t) = I(\omega, t) - I(\omega, -5\,{\rm ps})$. 
The energy dissipation occurs with two time scales having distinct 
pump-power dependences: at $t$\,$\lesssim$\,1\,ps, the energy 
dissipation rate positively depends on $p$, whereas at 
$t$\,$\gtrsim$\,1\,ps, it is independent of $p$. The fast dynamics 
at $t$\,$\lesssim$\,1\,ps is attributed to the net energy flow 
from the hot electrons to the hot COPs through electron-phonon scatterings, 
since this channel is increased quadratically with $p$ due to the 
increased populations of the hot electrons and hot COPs. 
Note that the scatterings among the hot electrons cannot account 
for the loss of the electronic energy \cite{Sarma}. 
At $t$\,$\sim$\,1\,ps, the electrons and COPs reach 
quasi-equilibrium ($T_{\it el}$\,=\,$T_{\it COP}$) after a sufficient 
number of scatterings, and the electron-COP composite 
dissipates its energy to the heat bath such as 
acoustic phonons \cite{MacDonald_AcousticCooling}, 
and hence the $p$-independent energy dissipation at $t$\,$\gtrsim$\,1\,ps.

\begin{figure}
\begin{center}
\includegraphics[width=87mm]{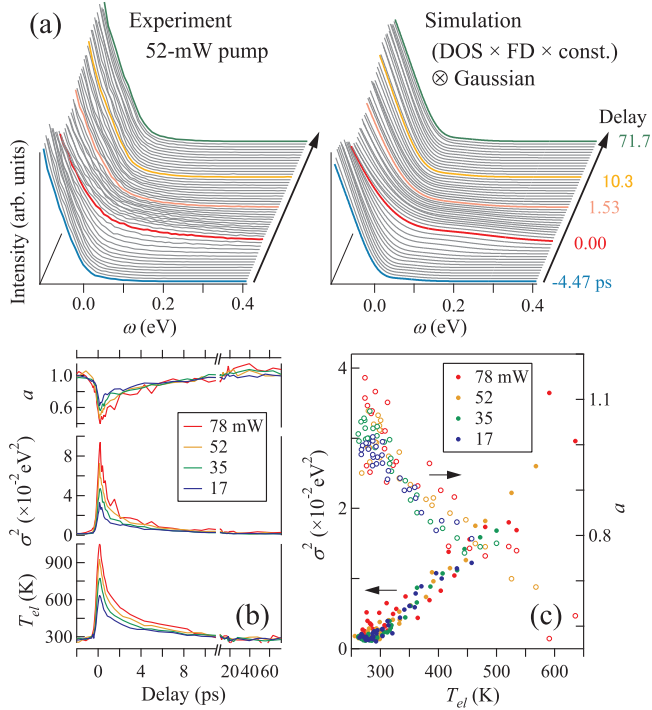}
\end{center}
\caption{\label{fig4}Optical-phonon broadenings during the transient.  
(a) TrPES spectra recorded under 52-mW pump (left) and the 
simulated spectra (right). 
(b) Fitting parameters. (c) A plot of the 
fitting parameters $\sigma^2$ and $a$ with respect to $T_{\it el}$ 
for $t$\,$>$\,1.0 ps. }
\end{figure}

Further analysis of the spectra also supports that the 
recovery dynamics at $t$\,$\gtrsim$\,1 ps is characterized 
by $T_{\it el}$\,=\,$T_{\it COP}$ and decay rates independent of $p$. 
First, we derive the decay-rate spectrum for 
the slower component 1/$\tau_2(\omega)$, which is obtained by 
fitting $I(\omega, t)$ at each energy with a double exponential function 
(Supplementary Information). As shown in Fig.\ \ref{fig3}(b), 
1/$\tau_2(\omega)$ is independent of $p$ and is quasi-linear to 
$\omega$ at $\omega$\,$\lesssim$\,0.3\,eV. Second, we quantify 
the spectral shape by simulating the spectrum as 
$I(\omega,t)$\,$=$\,$a(t)\int\,G(\omega-\omega',\sigma^2(t)+\sigma_R^2)
D(\omega')f(\omega', T_{\it el}(t))d\omega'$. 
Here, $G(\omega, \sigma^2+\sigma_R^2)$ is 
a Gaussian with FWHM of $\sqrt{\sigma^2+\sigma_R^2}$, $f(\omega, T_{\it el})$ 
is the Fermi-Dirac function, $a(t)$ is a scaling factor, 
and $D(\omega)$ is a DOS. The Gaussian broadening accounts for the 
spectral weight accumulating from lower energies, 
so that $\sigma^2$ 
becomes a measure of the number of the COPs in the transient. 
The spectra are nicely reproduced throughout the transient 
[Fig.\ \ref{fig4}(a) and Supplementary Information], 
and the fitting parameters are 
summarized in Fig.\ \ref{fig4}(b) and \ref{fig4}(c). 
One can see that $\sigma^2$ and $a$ for $t$\,$>$\,1.0\,ps scales with 
$T_{\it el}$ and does not explicitly depend on $p$. This indicates that 
the number of the COPs at $t$\,$>$\,1.0\,ps is a function of 
$T_{\it el}$, i.e., $T_{\it COP}$\,=\,$T_{\it el}$, 
and that the spectral shape at 
$t$\,$>$\,1.0\,ps is determined only by $T_{\it el}$, which is 
the temperature of the 
electron-COP composite.

\begin{figure}
\begin{center}
\includegraphics[width=87mm]{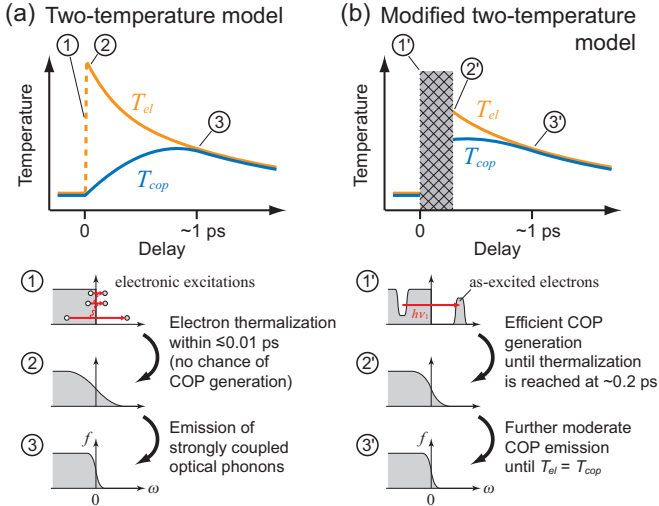}
\end{center}
\caption{\label{fig5}Two-temperature model (a) and beyond (b). 
The top panels show the transient change of 
$T_{\it el}$ and $T_{\it cop}$. Hatched area indicates 
temporal region where $T_{\it el}$ cannot be defined. 
The lower panels show electron distribution functions ($f$)
at some selected moments. The major difference in the 
COP-generation mechanism between the 
two is whether they are generated by the 
thermal electrons (a) or by the non-thermal electrons (b). }
\end{figure}

{\bf Discussion}.---
We experimentally find that the electronic distribution is 
non-thermal at $t \lesssim$ 0.2 ps, and also find evidence that 
the COP generation is mostly accomplished within this 
initial temporal region. 
This strongly indicates that the mechanism of 
the ultrafast COP 
generation is beyond the TTM scheme, see Fig.\ \ref{fig5}. 
Since the duration of the non-thermal electron distribution 
is longer than that in the TTM scheme, the high-energy electrons, 
which is considered to be favorable for generating high-energy 
COPs \cite{Milde_APL}, indeed have more chance to generate 
COPs before they degrade into low-energy thermalized electrons. 
Therefore, not only the largeness of the electron-phonon couplings but 
also the long durations of the non-thermal electron distribution 
may be crucial for understanding the 
ultrafast-and-efficient generation 
of the COPs in graphite. 
It is interesting to note that a break-down of TTM was also 
suggested in YBa$_2$Cu$_3$O$_{7-\delta}$ \cite{Pashkin}, 
in which the 
DOS becomes vanishingly small around $E_F$ in the 
$d$-wave superconducting state, similar to the case for graphite. 
Therefore, coupled dynamics directly involving 
the non-thermal electrons may be sought 
in materials that have vanishingly small DOS around $E_F$, 
for example in neutral graphene \cite{Novoselov}, 
in nodal superconductors \cite{Pashkin}, and on the 
surface of topological insulators \cite{Hasan_Rev}. 
Alternatively, we do not exclude the possibility that 
the hot COPs and hot electrons are 
co-generated by the pump, which may be viewed 
as a counterpart of the breakdown of adiabatic 
Born-Oppenheimer approximation \cite{BO}: 
in as much as the electrons cannot follow the motion 
of the lattice, COPs are simultaneously generated 
when the electronic excitation takes place. 
Whichever the case may be, our study reveals that there is 
a unique mechanism of ultrafast COP generation 
where the concept of temperatures is broken.

{\bf Methods}.---
The TrPES apparatus consists of 
an amplified Ti:sapphire laser system delivering 
$h\nu_1$\,=\,1.5 eV pulses 
of 170-fs duration with 250-kHz repetition and a 
hemispherical analyzer \cite{Kiss}, see Fig.\ \ref{fig1}(a). 
A portion of the laser is converted into $h\nu_4$\,=\,5.9 eV 
probing pulses 
using two non-linear crystals, $\beta$-BaB$_2$O$_4$ (BBO), 
and the time 
delay $t$ from the pump is controlled by a delay stage. 
The pump and probe pulses are $p$-polarized and have spot diameters 
of $\sim$0.8 and $\sim$0.3 mm, respectively, at the sample position. 
The intensity of the probe pulse is minimized to avoid 
space charge effects. 
Multi-photon photoemission due to the pump pulse 
is not observed in the dataset presented herein. 
$E_F$ is referenced to the Fermi cutoff of gold and the 
energy resolution $\sigma_R$ is 11\,meV. 
The base pressure of the photoemission chamber is 
1$\times$10$^{-10}$ Torr, and highly-oriented pyrolitic graphite 
is cleaved {\it in situ}.

\end{document}